\documentstyle[amssymb,aps,twocolumn]{revtex}

\begin{document}
\title{Testing the Kibble-Zurek Scenario \\
with Annular Josephson Tunnel Junctions}
\author{E.\ Kavoussanaki$^{a}$\thanks{{{{{{{{{{{{{{{{E-mail: {\tt %
e.kavoussanaki@ic.ac.uk}}}}}}}}}}}}}}}}}, R.\ Monaco$^{b}$\thanks{{{{{{{{{{{{%
{{{{E-mail: {\tt roberto@sa.infn.it}}}}}}}}}}}}}}}}} and R.\ J.\ Rivers$^{a}$%
\thanks{{{{{{{{{{{{{{{{E-mail: {\tt r.rivers@ic.ac.uk}}}}}}}}}}}}}}}}}}
\address{{\it a})Theoretical Physics, Blackett Laboratory, Imperial College, Prince\\
Consort Road, London, SW7 2BZ, U.K. \\
{\it b}) Istituto di Cibernetica del C.N.R., I-80072, Arco Felice (Na), Italy%
\\
and INFM-Dipartimento di Fisica, Universita' di Salerno, I-84081 Baronissi\\
(Sa), Italy}
\date{\today}
\maketitle

\begin{abstract}
In parallel with Kibble's description of the onset of phase transitions in
the early universe, Zurek has provided a simple picture for the onset of
phase transitions in condensed matter systems, supported by agreement with
experiments in $^{3}He$ and superconductors. In this letter we show how
experiments with annular Josephson tunnel Junctions can, and do, provide further support
for this scenario.
\end{abstract}


As the early universe cooled it underwent a series of spontaneous phase
transitions, whose potential inhomogeneities (monopoles, cosmic strings,
domain walls) have observable consequences, for structure formation in
particular. These defects appear because the correlation length $\xi $ of
the field (or fields) whose expectation value is the order parameter is
necessarily {\it finite} for a transition that is implemented in a finite
time

Using {\it nothing more} than simple causal arguments Kibble\cite
{kibble1,kibble2} made estimates of this early field ordering, and the
density of topological defects produced at GUT transitions at $10^{-35}s$.
Unfortunately, because the effects of their evolution are not visible until
the decoupling of the radiation and matter $10^{6}yrs$ later, it is
impossible to provide unambiguous checks of these predictions. However,
causality is such a fundamental notion that Zurek suggested\cite
{zurek1,zurek2} that identical causal arguments, with similar predictions, were applicable to condensed
matter systems for which direct experiments on defects could be performed.
The hope is that successful tests of these predictions could lead to a better understanding of
phase transitions in quantum fields.

Whether for the early universe or condensed matter, consider a quench of the
system in which its temperature $T(t)$ is reduced as time passes. In the
vicinity of the critical temperature $T_{c}$ we assume that the temperature $%
T$ decreases linearly with the time $t$ at a rate $dT/dt=-T_{c}/\tau _{Q}$, $%
\tau _{Q}$ being the quenching time.

Suppose that the 'equilibrium' correlation length $\xi _{eq}(t)=\xi
_{eq}(T(t)) $ of the order-parameter field, and its relaxation time $\tau (t)$,
diverge at $t=0$ (when $T=T_{c}$) as
\begin{equation}
\xi _{eq}(t)=\xi _{0}\bigg|\frac{t}{\tau _{Q}}\bigg|^{-\nu },\,\,\tau
(t)=\tau _{0}\bigg|\frac{t}{\tau _{Q}}\bigg|^{-\gamma }.  \label{csi+tau}
\end{equation}
\noindent The fundamental length and time scales $\xi _{0}$ and $%
\tau _{0}$  of a system are determined from its microscopic dynamics. One definition of $%
\tau (t)$ is that ${\bar{c}}(t)=\xi _{eq}(t)/\tau (t)$ denotes the maximum
speed, at time $t$, at which the order parameter can change. In quantum
field theory ${\bar{c}}(t)=c_{0}$, the speed of light in vacuo.

Although $\xi _{eq}(t)$ diverges at $t=0$ this is not the case for the true
non-equilibrium correlation length $\xi (t)$.
Kibble and Zurek made two assumptions. Firstly, the
correlation length ${\bar{\xi}}$ of the fields that characterizes the onset
of order is the equilibrium correlation length ${\bar{\xi}}=\xi _{eq}({\bar{t%
}})$ at some time ${\bar{t}}$ constrained by causality. Secondly, at this
time, defects appear with separation $\xi _{def}=O({\bar{\xi}})$.

To determine ${\bar{t}}$ we rephrase the original Kibble-Zurek argument in a
way appropriate to Josephson tunnelling Junctions (JTJs), so as only to discuss times $t>0$. We begin by
noting that, in the adiabatic regime away from the transition, static defects can
be thought of as kinks, balls, lines\footnote{%
For $^{3}He$ vortices can be more complicated, but our general argument is
unaffected.} or sheets of 'false' vacuum or disordered ground state, of
thickness $O(\xi _{eq}(t))$. Thus ${\dot{\xi}}_{eq}(t)=d\xi _{eq}(t)/dt<0$
measures the rate at which these defects contract, i.e., the speed of
interfaces between ordered and disordered ground states. Since ${\dot{\xi}}%
_{eq}(t)$ decreases with time $t>0$, the {\it earliest} possible time $t$ at
which defects could possibly appear is determined by $|{\dot{\xi}}_{eq}(t)|={%
\bar{c}}(t)$, given our definition of ${\bar{c}}(t)$. Although this gives a
{\it lower} bound for ${\bar{t}}$, as an order of magnitude estimate we identify
this time $t$ with ${\bar{t}}$, whence
\begin{equation}
\tau _{0}\,\ll \,\,{\bar{t}}=(\tau _{Q}^{\gamma }\tau _{0})^{1/(\gamma
+1)}\ll \tau _{Q}.  \label{tbar}
\end{equation}
\noindent The corresponding {\it smallest}\cite{zurek2} correlation length is
\begin{equation}
{\bar{\xi}}=\xi _{eq}({\bar{t}})=\xi _{0}\bigg(\frac{\tau _{Q}}{\tau _{0}}%
\bigg)^{\nu /(\gamma +1)}\gg \xi _{0}.  \label{xiZ}
\end{equation}
\noindent Because of the qualitative nature of the arguments, factors close to unity
are omitted\footnote{%
These results, without any additional factors, were originally obtained by Zurek on considering
the time $-\overline{t}$ at which the field freezes in.}. Further, at the
same level of approximation, we shall use mean field critical indices
throughout. Measurements\cite{helsinki,grenoble} of {\it total} vortex
density in transitions of $^{3}He-B$ support the result (\ref{xiZ}), when
taken together with $\xi _{def}=O({\bar{\xi}})$.

As an independent test of the assumptions Zurek suggested using (\ref{xiZ})
to measure {\it topological} defect density (in which defects and
anti-defects carry opposite weight). This is most easily done in 'one-dimensional' annular
geometries, for which experiments were originally proposed \cite{zurek2}
with $^{4}He$ which, however, has ${\bar{\xi}}$ so small that the creation
of an effectively  one-dimensional system is extremely difficult. A recent experiment \cite
{carmi}with annular arrays of high-$T_{c}$ superconducting islands coupled
by grain boundary Josephson junctions confirms part of the picture, but
suffers in enforcing a predetermined domain structure. We also wish to check
prediction (\ref{xiZ}) by using annular JTJs. As we shall see, for such JTJs ${%
\bar{\xi}}$ is {\it macroscopically }large, permitting them to be
effective one-dimensional systems.

An annular JTJ consists of two superimposed annuli of ordinary
superconductors of thickness $d_{s}$, separated by a layer of oxide of
thickness $d_{ox}$, whose relative dielectric constant is $\epsilon _{r}$.
Its order-parameter is the relative phase angle $\phi =\theta _{1}-\theta
_{2}$ of the complex order parameters $\Psi _{1}=\rho _{1}\exp (i\theta
_{1}) $ and $\Psi _{2}=\rho _{2}\exp (i\theta _{2})$ of the two
superconductors (labelled $1$ and $2$). After the transition has been
implemented, in the adiabatic regime at temperature $T$, $\phi $ satisfies the dissipative,
one-dimensional sine-Gordon (SG) equation
\begin{equation}
\frac{\partial ^{2}\phi }{\partial x^{2}}-\frac{1}{{\bar{c}}^{2}(T)}\frac{%
\partial ^{2}\phi }{\partial t^{2}}-\frac{b}{{\bar{c}}^{2}(T)}\frac{\partial
\phi }{\partial t}=\frac{1}{\lambda _{J}^{2}(T)}\sin \phi ,  \label{sine-gordon}
\end{equation}
\noindent with periodic boundary conditions\cite{Lomdahl}; $x$ measures the
distance along the annulus, its width $w\ll \lambda _{J}(T)$ being ignored and $%
b$ is a characteristic frequency that accounts for the viscous drag.
The velocity ${\bar{c}=\bar{c}%
(T)}$, which depends on the nature of the junction, is the Swihart\cite
{Swihart} velocity, the speed of light in a
superconducting-insulating-superconducting transmission line. In the
Josephson context, it determines the maximum speed at which the order
parameter $\phi $ can change.

The topological defects of the JTJ, the solitons of the
sine-Gordon theory, are termed {\it fluxons}. Their static equilibrium thickness is
the Josephson coherence length $\lambda _{J}(T)$, which
plays the role of $\xi _{eq}(T)$ earlier.

Let us attempt to repeat the Kibble-Zurek analysis directly on quenching a JTJ
with quench time $\tau _{Q}$. For simplicity we begin with a symmetric JTJ,
in which the electrodes are made of identical materials with common critical
temperatures $T_{c}$. At time $t$ after the transition $\lambda _{J}(t)=\lambda _{J}(T(t))$ is
given by
\begin{equation}
\lambda _{J}(t)=\sqrt{\hbar /2e\mu _{0}d_{e}(t)J_{c}(t)}.  \label{JJlength}
\end{equation}
\noindent in which $J_{c}$ is the critical Josephson current density.
In (\ref{JJlength}) $d_{e}(t)$ is the magnetic thickness. Specifically, if $%
\lambda _{L}(t)$ is the London penetration depth of the two (identical)
superconducting sheets, then
\[
d_{e}(t)=d_{ox}+2\lambda _{L}(t)\tanh \frac{d_{s}}{2\lambda _{L}(t)},
\]
where $\lambda _{L}(t)=\lambda _{L}(0)/\sqrt{1-(T(t)/T_{c})^{4}}%
\simeq \lambda _{L}(0)/2\sqrt{{\bar{t}}/\tau _{Q}}$.  Neglecting
the barrier thickness $d_{ox}\ll d_{s},\,\lambda _{L}$ gives $d_{e}=d_{s}$ close to $T_{c}$%
, i.e., the magnetic thickness equals the film thickness and can be set
constant in (\ref{JJlength}).

All the $t$-dependence of $\lambda _{J}$ resides in $J_{c}$ which, for the
symmetric JTJ has the form\cite{Barone}
\begin{equation}
J_{c}(t)=\frac{\pi }{2}\frac{\Delta (t)}{e\rho _{N}}\tanh \frac{\Delta (t)}{%
2k_{B}T(t)}.  \label{Jc}
\end{equation}
\noindent In (\ref{Jc}) $\Delta (t)$ is the superconducting gap energy and
varies steeply near $T_{c}$ as
\[
\Delta (t)\simeq 1.8\,\Delta (0)\bigg(1-\frac{T(t)}{T_{c}}\bigg)%
^{1/2}=1.8\,\Delta (0)\sqrt{\frac{t}{\tau _{Q}}},
\]
\noindent and $\rho _{N}$ is JTJ normal resistance per unit area.
Introducing the dimensionless quantity $\alpha =1.6\Delta (0)/k_{B}T_{C}$
whose typical value\footnote{$\Delta (0)$ and $J_{c}(0)$ denote the
respective values at $T=0.$} is between 3 and 5, enables us to write $%
J_{c}(t)$ as
\begin{equation}
J_{c}(t)\simeq \alpha J_{c}(0)\bigg(1-\frac{T(t)}{T_{c}}\bigg) =\alpha
J_{c}(0)\frac{t}{\tau _{Q}}.  \label{Jct}
\end{equation}
\noindent Thus, in the vicinity of the transition,
\begin{equation}
\lambda _{J}(t)=\xi _{0}\bigg(\frac{\tau _{Q}}{t}\bigg)^{1/2},  \label{xeq1}
\end{equation}
\noindent corresponding to $\nu =1/2$ in (\ref{csi+tau}), where
\begin{equation}
\xi _{0}=\sqrt{\frac{\hbar }{2e\mu _{0}d_{s}\alpha J_{c}(0)}}.
\label{csi_not}
\end{equation}
On the other hand, for a finite electrode thickness tunnel junction, the
Swihart velocity takes the form\cite{Barone}

\[
{\bar{c}(t)}=c_{0}\sqrt{d_{ox}/\epsilon _{r}d_{i}(t)},
\]
where
\[
d_{i}(t)=d_{ox}+2\lambda _{L}(t)\coth \frac{d_{s}}{2\lambda _{L}(t)}\simeq
\frac{\lambda _{L}^{2}(0)}{d_{s}}%
{\tau _{Q} \overwithdelims() t}%
,
\]
near the transition. Thus $\overline{c}(t)$ shows critical slowing down at
the transition, as
\[
{\bar{c}(t)=\bar{c}}_{0}\left( \frac{t}{\tau _{Q}}\right) ^{1/2},
\]
where ${\bar{c}}_{0}=c_{0}\sqrt{d_{s}d_{ox}/\epsilon _{r}\lambda _{L}^{2}(0)}$.
These indices $(\nu = 1/2,\gamma =1)$
are typical of condensed matter systems. The causal constraint
gives ${\bar{t}}=\sqrt{\tau _{0}\tau _{Q}}$,
with $\tau _{0}=\xi _{0}/{\bar{c}}_0$. Inserting reasonable values%
\cite{Barone} of $\xi _{0}=10\,\mu m$ and ${\bar{c}}_0=10^{7}\,m/s$, gives $%
\tau _{0}=1\,ps$, and assuming $\tau _{Q}=1\,s$, we find ${\bar{t}}\simeq
1\mu s$.
The causal Josephson
penetration length is then
\begin{equation}
\overline{\lambda }_{J}=\lambda _{J}(\overline{t})=\xi _{0}\bigg(\frac{\tau
_{Q}}{\overline{t}}\bigg)^{\frac{1}{2}}=\xi _{0}\bigg(\frac{\tau _{Q}}{\tau
_{0}}\bigg)^{\frac{1}{4}}= 10\,mm.  \label{csi}
\end{equation}


This $\overline{\lambda }_{J}$, which should characterise fluxon separation at a
quench for a symmetric JTJ is far too large. Fortunately, the manufacture of
JTJs typically yields {\it non-symmetric} devices with more acceptable
properties. Suppose the two superconductors, $%
1$ and $2$, now have different critical temperatures $T_{c2}>T_{c1}$. Fluxons
only appear at temperatures $T<T_{c1}$, from which we measure our time $t$.
At this time
\[
\Delta _{2}(T_{c1})\simeq 1.8\,\Delta _{2}(0)\bigg(1-\frac{T_{c1}}{T_{c2}}%
\bigg)^{1/2},
\]
\noindent and $\Delta _{1}(t)\simeq 1.8\,\Delta _{1}(0)\sqrt{t/\tau _{Q}}$.
The critical Josephson current density $J_{c}^{\prime }(t)$ for a
non-symmetric JTJ, being proportional to $\Delta _{1}(t)\Delta _{2}(t)$ \cite
{Barone}, behaves just after the transition as
\begin{equation}
J_{c}^{\prime }(t)\approx \bigg(1-\frac{T_{c1}}{T_{c2}}\bigg)^{1/2}\alpha
^{\prime }J_{c}^{\prime }(0)\bigg(\frac{t}{\tau _{Q}}\bigg)^{1/2},
\label{Jct2}
\end{equation}
\noindent where $J_{c}^{\prime }(0)=\pi \Delta _{1}(0)\Delta _{2}(0)/[\Delta
_{1}(0)+\Delta _{2}(0)]e\rho _{N}$, and $\alpha ^{\prime }=[\Delta
_{1}(0)+\Delta _{2}(0)]/k_{B}T_{c,1}$, provided $\Delta _{2}(T_{c,1})\ll
2\pi k_{B}T_{c,1}$. This is the case here.

\noindent The crucial difference between (\ref{Jct2}) and (\ref{Jct}) is in
the critical index. Near $t=0$, we now find
\begin{equation}
\lambda _{J}(t)=\xi _{0}\bigg(1-\frac{T_{c1}}{T_{c2}}\bigg)^{-1/4}\bigg(%
\frac{\tau _{Q}}{t}\bigg)^{1/4},  \label{xeq2}
\end{equation}

\noindent where $\xi _{0}$ is as in (\ref{xeq1}), since $J_{c}^{\prime }(0)$
is indistinguishable from $J_{c}(0)$ and $\alpha ^{\prime }$ is comparable
to $\alpha $. For the critical behavior (\ref{xeq2}) to be valid, rather than (%
\ref{xeq1}) we need
$1 -T_{c1}/T_{c2}\gg O(\overline{t}/\tau _{Q})%
=O(10^{-6})$, which is always the case.
For a typical value $\left( 1-T_{c1}/T_{c2}\right) =0.02$
the critical time $\overline{t}$ is now determined by ($\gamma = 3/4$)

\[
{\bar{t}}=\tau _{0}^{4/7}\tau _{Q}^{3/7}\bigg(1-\frac{T_{c1}}{T_{c2}}\bigg)^{-1/7}\simeq 0.24\mu s,
\]
with our parameters.  In turn,
\begin{equation}
\lambda _{J}(\overline{t})\simeq \xi _{0}\bigg(1-%
\frac{T_{c1}}{T_{c2}}\bigg)^{-1/4}\bigg(\frac{\tau _{Q}}{\tau _{0}}\bigg)%
^{1/7}\simeq 1.4mm
\label{xeq4}
\end{equation}
is an
order of magnitude smaller than ${\bar\lambda}_J$ of (\ref{csi}).

While new experiments are required, old experiments on JTJs by one of us\cite
{Roberto} are compatible with these predictions, although their specific parameters are not
optimal. In these experiments non-symmetric
annular $Nb/Al-AlOx/Nb$ JTJs ($T_{c,2}/T_{c,1}-1\approx 0.02$) with
circumference $C=0.5\,mm$ were quenched with a quench time $\tau _{Q}=O(1s)$%
. The intention was, primarily, to produce fluxons for further experiments,
and the density at which they were produced was secondary. From the
parameters quoted in \cite{Roberto} for sample B, we estimated $\xi
_{0}\simeq 6.5\,\mu m$, ${\bar{c}}_{0}\simeq 10^{7}\,m/s$ and $\tau _{0}\simeq
0.65ps$. Inserting these specific values in (\ref{xeq4}) gives ${\bar{\lambda}_J}\simeq 1mm
$ (with experimental uncertainty of up to $50\%$). Although $C\simeq{\bar{\lambda}_J} $
we would have expected to see a fluxon a
few percent of the time, given that the variance $\Delta n$ in the number of fluxons is
$\Delta\phi /2\pi$. Indeed, in practice (invariably single) defects formed
once every 10-20 times.

We have no detailed knowledge of how the cooling takes place,
but do not expect temperature inhomogeneities to be important.
The critical slowing down of ${\bar{c%
}(t)}$ provides a necessary condition for defects to survive inhomogeneity\cite
{volovik} and, with empirically comparable $\xi
_{0}$, $\tau _{0}$, and $\tau _{Q}$, the situation is no better or worse
for JTJs than with any other superconducting system undergoing a mechanical quench.
Other samples of the same circumference but with different ${\bar{\lambda}_J}$
have been tested. Although none had $C/{\bar{\lambda}_J}$ large,
it was observed that the likelihood of seeing a
fluxon was greater the larger its value, as we would have predicted,
although this was not quantified.
This suggests that temperature inhomogeneities are not the direct
cause of the observed fluxons.

There are theoretical, as well as experimental, uncertainties.
The SG equation (\ref{sine-gordon}) can only make sense once
the individual superconductors
have adjusted themselves.  Repeating Zurek's analysis
of the Gross-Pitiaevsky equation\footnote{Justified here by the success of
Feynman's coupled model equations for $\Psi_1$ and $\Psi_2$. For example, see Ref.10.}
for individual superconductors\cite{zurek2}
 gives a minimum time at which the sine-Gordon equation is valid of
${\bar{t}}_{S}=\sqrt{\overline{\tau }_{0}\tau _{Q}}$
where $\overline{\tau }_{0}$ in (\ref{csi_not}) is now determined
\cite{zurek2} from Gorkov's equation as
$\overline{\tau }_{0}=\pi \hbar/16k_{B}T_{c}\approx 0.15ps$
for $T_{c}\approx 10\,K$. The resulting ${\bar{t}}_{S}\approx 0.4 \mu s$
is commensurate with the values of ${\bar{t}}$ for the typical symmetric and non-symmetric cases,
falling between them.
Whereas this suggests that the SG equation is valid for symmetric JTJs at time ${\bar{t}}$, it also
suggests that we should evaluate $\lambda_J(t)$ at ${\bar t}_S$, rather than ${\bar t}$
for the non-symmetric case.
However, for our typical parameter values the difference between $\lambda _{J}(\overline{t})$ and
\begin{equation}
\lambda _{J}(\overline{t}_S)\simeq \xi _{0}\bigg(1-%
\frac{T_{c1}}{T_{c2}}\bigg)^{-1/4}\bigg(\frac{\tau _{Q}}{{\bar\tau} _{0}}\bigg)%
^{1/8}\simeq 1.1mm  \label{xeq3}
\end{equation}
is so small as to be ignorable, given the crudity of the bounds.  For the specific sample B of
 \cite{Roberto} the decrease is similar, at $\lambda _{J}(\overline{t}_S)\simeq 0.7mm$, and equally ignorable.

Further, although the prediction (\ref
{xiZ}), together with $\xi _{def}=O({\bar{\xi}})$, has been taken, without
additional qualification, as the direct basis for the successful experiments\cite
{helsinki,grenoble} in $^{3}He$, and experiments in $^{4}He$ \cite{lancaster,lancaster2} and
high-$T_{c}$ superconductors\cite{technion}, the causality argument that we
have presented here is very simplistic. For superfluids obeying
time-dependent Ginzburg-Landau (TDGL) theory (and QFT) we know\cite{RKK}
that, at early times, the length ${\bar{\xi}}$ is, correctly, the
correlation length of the fields when they have frozen in after the
transition. However, we also know\cite{halperin,maz} that the separation of
defects is determined largely by the separation of the zeroes of the fields
which define their cores. The separation of zeroes is a function of the
{\it short-range} behavior of the correlation functions\cite{halperin,maz}, rather
than the long-range behavior that determines ${\bar{\xi}}$. A priori, ${\bar{%
\xi}}$ does not characterize defect separation.

Nonetheless,
several numerical\cite{zurek3,calzetta} and analytic calculations\cite
{RKK,ray,ray2}, based on TDGL theory, have confirmed that the critical index
of (\ref{xiZ}) is, indeed, the correct behavior for defect separation. The
reason why this is so is essentially a matter of dimensional analysis.
The density of zeroes is, approximately, a {\it ratio} of moments of the
power in the field fluctuations, at early times at least. This leads to
strong cancellations of the effects of the microscopic interactions
of the system in question.

At the same time, the
critical time $\overline{t}$ characteristically underestimates the time at
which the order parameter achieves its equilibrium magnitude, which is a
more sensible time to begin to count defects. However, if these other systems are a guide\cite{RKK,ray,ray2}
the true time $t^{*}$is $t^{*}=O({\bar{t}})$, since the unstable long
wavelength modes that set up large scale ordering have amplitudes that grow
exponentially. As a result any new scales only occur logarithmically in $%
t^{*}/{\bar{t}}$. In fact, a limited
calculation, with $^{4}He$ in mind, suggests\cite{edik} that
$\xi _{def}(t^{*})\approx {\bar{\xi}}$  when
counting topological density on an annulus. Thus, although the causality bounds
are not saturated, their consequences (\ref{tbar}) and (\ref{xiZ}) survive
qualitatively and justify experimental
confirmation.

That causality is now seen as a constraint, but not the microscopic mechanism,
helps explain why the most recent $^{4}He$
experiment\cite{lancaster2} failed to see any vortices. A major reason
(nothing to do with causality) is
that vortices are most likely to decay much more rapidly\cite{ray2} than
they were originally thought to do. However, because the $^{4}He$ quenches take
place entirely within the Ginzburg regime, thermal fluctuations make
individual defects scale dependent\cite{ray,ray2}, and simple dimensional
analysis most likely breaks down, in a way that causality would not have suggested.
This is not the case with superconductors, for which the Ginzburg regime is very small.

For that and other reasons, we are sufficiently optimistic to be
currently examining the feasability of fabricating JTJs with larger
values of $C/{\bar{\lambda}_J}$ with which to perform new experiments.

E.K. and R.R. thank the EU Erasmus/Socrates programme for financial support
and the University of Salerno for hospitality. This work is the result of a
network supported by the European Science Foundation.

\end{document}